\documentclass[11pt]{article}

\usepackage{arxiv}

\usepackage{amsmath, amssymb, amsthm}
\usepackage{booktabs}
\usepackage{longtable}
\usepackage{array}
\usepackage{threeparttable}
\usepackage{graphicx}
\usepackage{adjustbox}
\usepackage{tabularx}
\usepackage{microtype}
\usepackage{nohyperref}
\usepackage[nameinlink,noabbrev]{cleveref}
\usepackage{enumitem}
\usepackage{caption}
\usepackage{subcaption}
\usepackage{tikz}
\usetikzlibrary{arrows.meta,positioning,shapes.geometric}
\usepackage[authoryear,round]{natbib}
\usepackage{float}

\renewcommand{\arraystretch}{1.12}

\title{AI-Assisted Economic Measurement from Survey Instruments:\\
Evidence from Public-Employee Pension Choice}

\author{
  Tiancheng Wang \\
  Hoover Institution, Stanford University \\
  \texttt{tpw5237@stanford.edu} \\
  \And
  Krishna Sharma \\
  Hoover Institution, Stanford University \\
  \texttt{ksharma3@stanford.edu}
}

\date{}

\newcommand{\R}{\mathbb{R}}

\begin{document}
\maketitle

\begin{abstract}
We develop an iterative framework for economic measurement that leverages large language models to extract measurement structure directly from survey instruments. The approach maps survey items to a sparse distribution over latent constructs through what we term a ``soft mapping,'' aggregates harmonized responses into respondent-level subdimension scores, and disciplines the resulting taxonomy through out-of-sample incremental validity tests and discriminant validity diagnostics. The framework explicitly integrates iteration into the measurement construction process. Overlap and redundancy diagnostics trigger targeted taxonomy refinement and constrained remapping, ensuring that added measurement flexibility is retained only when it delivers stable out-of-sample performance gains. Applied to a large-scale public employee retirement plan survey, the framework identifies which semantic components contain behavioral signal and clarifies the economic mechanisms, such as beliefs versus constraints, that matter for retirement choices. The methodology provides a portable measurement audit of survey instruments that can guide both empirical analysis and survey design.
\end{abstract}


\section{Introduction}
\label{sec:intro}

Survey instruments remain central to economic measurement. They operationalize latent constructs, including beliefs, expectations, constraints, preferences, and knowledge, through structured questionnaires. By compressing complex economic concepts into concise item stems, researchers can field large-scale instruments and collect rich data on individual behavior and outcomes. This compression, however, creates a fundamental measurement challenge. Many survey items bundle multiple mechanisms within a single question, so any strict one-to-one assignment of items to constructs can be problematic. When an item reflects both beliefs and knowledge, or both constraints and preferences, the resulting measurement scores can blur conceptual boundaries and mix distinct economic mechanisms. We call this phenomenon \emph{construct contamination}, and it has important consequences for empirical analysis. Contaminated constructs are harder to interpret, less portable across contexts, and they complicate the design of targeted policy interventions \citep{cronbach1955construct, campbell1959convergent}.

Recent developments in large language models \citep{openai2023gpt, bommasani2021opportunities} create new opportunities to address this measurement challenge. These models can parse item semantics systematically at scale and produce replicable, text-based classifications of survey content. A growing literature in psychology and measurement science shows that embeddings and LLMs can quantify semantic overlap across items, diagnose taxonomic inconsistencies, and recover questionnaire structure from text alone \citep{boke2025observer, huang2025embedding, kambeitz2025empirical, wulff2025semantic, hommel2025language}. This work provides a methodological foundation that we build upon. The language of a survey instrument contains actionable information about its measurement structure, and LLMs offer scalable tools to extract that structure in a transparent and auditable manner.

For economics, the central question is not whether LLMs can classify survey items, but whether such classifications improve economic measurement in ways that matter for empirical analysis and policy. For an LLM-based taxonomy to be useful, it must deliver clearer separation of economic mechanisms and more informative heterogeneity in outcomes, and these improvements must be stable out of sample \citep{athey2019machine, mullainathan2017machine}. Simply expanding the number of constructs can mechanically improve in-sample fit without enhancing measurement quality, and LLM-generated mappings can introduce new forms of overlap and contamination. A credible framework for AI-assisted economic measurement therefore requires coupling scalable measurement construction with rigorous econometric validation \citep{chernozhukov2018double}. We evaluate the value of an LLM-proposed taxonomy by the same standards applied to any measurement system. Does it improve our ability to predict outcomes, does it sharpen our understanding of mechanisms, and do these gains replicate beyond the sample used to construct the measures \citep{kleinberg2015prediction}?

We propose an iterative measurement framework that treats the survey instrument itself as data and uses econometric diagnostics to discipline LLM-generated measurement proposals. Starting from item stems, an LLM induces a multidimensional taxonomy of subdimensions and produces sparse item-to-subdimension weight vectors, adapting methods from \citet{kargupta2025taxoadapt}. We interpret these weights as a soft mapping in which each item loads on a simplex over subdimensions, which makes cross-loading explicit and auditable. We then combine this mapping with harmonized survey responses to construct respondent-level scores for each subdimension. Critically, we do not treat the LLM output as ground truth. Instead, we evaluate each proposed construct using incremental out-of-sample validity tests and discriminant validity diagnostics designed to detect construct overlap \citep{borsboom2004concept}. When diagnostics indicate contamination or instability, we apply targeted refinement operators that modify the taxonomy or tighten mapping constraints. We iterate until overlap diagnostics stabilize and out-of-sample gains plateau. In this way, the framework operationalizes a discipline mechanism in which added measurement flexibility is retained only when justified by stable empirical performance.

We illustrate the framework using a large-scale survey of public employees on their valuation of defined benefit pension plans relative to defined contribution alternatives, originally analyzed by \citet{giesecke2022much}. Our out-of-sample diagnostics show that only a small subset of measurement subdimensions delivers stable incremental information for predicting retirement plan choices. The strongest signal is concentrated in mechanisms related to tenure and career-stage lock-in, which improves prediction for both acceptance of a hypothetical plan switch and the required employer contribution rate. Financial literacy content contributes meaningfully to predicting acceptance but provides minimal incremental value for predicting required contribution rates, a pattern consistent with the original study's findings but now validated through systematic out-of-sample testing. The diagnostics also identify construct overlap between literacy and perceived plan generosity, which motivates a targeted refinement that separates beliefs about plan value from objective financial knowledge. This refinement improves discriminant validity without inflating the taxonomy unnecessarily. Overall, the exercise shows how our framework can convert a high-dimensional survey instrument into a validated set of economic measurements, clarify which mechanisms matter for specific outcomes, and guide improvements in survey design.

The remainder of the paper proceeds as follows. Sections~2--5 develop a general framework for AI-assisted economic measurement and its validation. Sections~6--7 apply the framework to a large public-employee pension survey. Sections~8--9 discuss implications and conclude.

\section{Related Literature and Conceptual Framework} \label{sec:literature}
Our contribution lies at the intersection of economic measurement, machine learning, and text-as-data methods. We situate the framework within three closely related literatures: semantic measurement and text-based approaches, iterative taxonomy construction, and econometric validation of machine learning--assisted measurement. We also place the contribution in the context of the emerging literature on applications of artificial intelligence in economic research

\subsection{Text as data and semantic measurement}

The use of text-based methods in economics has expanded rapidly over the past decade. \citet{gentzkow2019text} and \citet{grimmer2022text} provide comprehensive overviews of text-as-data methods in the social sciences, emphasizing the value of treating language as a source of structured information. \citet{benoit2020text} offers a complementary perspective focused on the methodological challenges of converting text into structured data, while \citet{ash2023text} survey text algorithms in economics with applications spanning policy analysis, financial markets, and organizational behavior. Representative applications include \citet{kelly2021text}, who develop methods for text selection in economic prediction, \citet{hansen2018transparency}, who analyze Federal Reserve communications using computational linguistics, and \citet{baker2016measuring}, who construct indices of economic policy uncertainty from news text.

Our approach extends this literature by applying semantic analysis not to naturally occurring text corpora, but to the controlled language of survey instruments, where each item is deliberately designed to measure an underlying construct. While existing work demonstrates that item language contains recoverable structure, it does not address how such structure should be disciplined using econometric criteria.

Recent work in psychology and measurement science shows that embeddings and large language models can recover psychometric structure from item text alone. \citet{wulff2025semantic} demonstrate that semantic embeddings reveal taxonomic inconsistencies in psychological measurement and can guide construct refinement. \citet{boke2025observer} use LLMs to assess content overlap in mental health questionnaires, providing observer-independent diagnostics of item redundancy. \citet{hommel2025language} show that language models can infer correlations between psychological items and scales from text alone, while \citet{huang2025embedding} and \citet{kambeitz2025empirical} provide additional evidence that LLMs capture meaningful structure in psychological constructs and identify redundancy in existing instruments.

These studies establish that the language of measurement instruments is informative about construct boundaries and empirical relationships. We adopt this premise but impose a stricter evaluation standard. Rather than treating semantic structure as an endpoint, we treat LLM-generated taxonomies as measurement proposals that must be validated through econometric performance. A taxonomy is retained only if it improves out-of-sample prediction and passes diagnostics designed to detect construct contamination. Our framework adapts this logic to survey instruments, replacing corpus-coverage objectives with out-of-sample economic validation.

\subsection{Iterative taxonomy construction}

The iterative refinement component of our framework draws on recent work in natural language processing on adaptive taxonomy construction. \citet{kargupta2025taxoadapt} develop TaxoAdapt, an LLM-based procedure for constructing and refining multidimensional taxonomies aligned with evolving research corpora through repeated classification, expansion, and consolidation. We adapt this logic to the finite corpus of survey items, treating the instrument itself as the target corpus. Crucially, we replace corpus-coverage criteria with econometric diagnostics to govern iteration.

The key distinction is that we do not expand the taxonomy to maximize semantic coverage. Instead, we refine it to minimize construct overlap while maximizing out-of-sample predictive value. Iteration is therefore disciplined by empirical performance rather than by textual completeness.

\subsection{Machine learning and econometric validation}

Our emphasis on out-of-sample validation connects to a broad literature at the intersection of machine learning and econometrics. \citet{mullainathan2017machine} and \citet{athey2019machine} provide foundational frameworks for incorporating machine learning into economic research, emphasizing prediction tasks, cross-validation, and the avoidance of overfitting. \citet{abadie2020sampling} clarify the distinction between sampling-based and design-based inference in regression analysis, a distinction that is relevant for interpreting out-of-sample performance in survey contexts. \citet{kleinberg2015prediction} distinguish prediction policy problems, where forecast accuracy is central, from causal inference problems, where estimating treatment effects is the goal.

Our framework addresses a measurement problem that lies between these categories. We use predictive performance as a diagnostic for measurement quality, while the ultimate objective is to construct portable measures of economic mechanisms that can support causal inference and policy analysis. Importantly, we apply these principles to measurement construction itself rather than only to downstream estimation.

The validation protocol we develop draws on established practices in machine learning for model selection and performance evaluation. We use embedded cross-validation to guard against post-selection bias induced by iterative refinement, following principles in \citet{chernozhukov2018double} and \citet{athey2025surrogate}. The incremental validity tests we apply are analogous to methods used in psychology for construct validation \citep{cronbach1955construct, campbell1959convergent}, but they are adapted to an out-of-sample prediction context. By requiring that each proposed construct improve predictive performance in held-out data, we impose a discipline that prevents the taxonomy from expanding to fit noise.

\subsection{Construct validity and discriminant validity}

The discriminant validity diagnostics we employ are grounded in classic psychometric theory. \citet{campbell1959convergent} introduce the multitrait-multimethod matrix as a tool for assessing convergent and discriminant validity, arguing that constructs should correlate with related measures while remaining empirically distinct from conceptually different constructs. \citet{belch1977discriminant} provide early applications of discriminant validity assessment in consumer research. \citet{cronbach1955construct} and \citet{borsboom2004concept} emphasize that construct validity is an ongoing process of empirical testing rather than a one-time certification.

Our framework operationalizes these principles in a machine learning context by treating high correlations among purportedly distinct constructs as diagnostic of contamination. We use conditional contribution tests to determine whether overlap-prone constructs add unique information beyond existing measures.

\subsection{AI applications in economic research}

Our work also contributes to a growing literature on applications of artificial intelligence in economics beyond conventional prediction tasks. \citet{varian2018artificial} discusses how artificial intelligence is reshaping industrial organization and economic analysis, while \citet{einav2014economics} examine how big data is transforming empirical economics. \citet{brynjolfsson2021productivity} document complementarities between general-purpose technologies such as AI and organizational intangibles, highlighting the importance of measurement systems capable of capturing evolving relationships. \citet{hartley2024labor} study the labor market effects of generative AI, underscoring the need for robust measurement frameworks as AI adoption accelerates.

Related work examines the use of language models in survey and labor market contexts. \citet{argyle2023out} explore the use of language models to simulate human survey responses, raising questions about validation and bias. \citet{horton2017effects} show that algorithmic recommendations can affect labor market matching and outcomes. \citet{raji2020closing} develop frameworks for algorithmic auditing and accountability, emphasizing transparency and validation in deployed AI systems. Our application differs from this literature in that we use LLMs not to replace human respondents or generate recommendations, but to audit and improve measurement systems. The framework provides a disciplined methodology for leveraging AI to enhance data quality and measurement transparency.

Relative to this literature, our contribution is threefold. First, we develop an instrument-level measurement framework that maps survey items to a sparse distribution over subdimensions, making cross-loading explicit and auditable. Because the mapping is defined at the instrument level rather than the sample level, it is portable across datasets that field the same or similar items. Second, we integrate econometric diagnostics directly into the measurement construction process, using incremental out-of-sample performance and stability to determine which constructs to retain, refine, or discard. Third, we operationalize a disciplined refinement loop in which overlap diagnostics trigger targeted taxonomy modifications and constraint tightening, ensuring that added flexibility is used only when it delivers stable empirical gains. The result is a transparent and replicable procedure for converting survey instruments into validated economic measurements.

\section{Measurement Framework}
\label{sec:framework}

This section defines the measurement framework abstractly, independent of any particular dataset. We describe the objects, the LLM-based proposal mechanism, and how these elements fit into an iterative measurement loop.

\subsection{Objects and notation}
Let survey items (question stems) be indexed by $j=1,\dots,J$ and respondents by $i=1,\dots,N$. Each item $j$ has a fixed stem text $q_j$ (and, where relevant, answer options). Let $x_{ij}\in\R$ denote respondent $i$'s harmonized response to item $j$ (possibly missing). The goal is to measure a set of latent subdimensions indexed by $k=1,\dots,K$ organized in a multidimensional taxonomy.

We construct two objects: an item-to-subdimension weight matrix $W=\{w_{jk}\}\in[0,1]^{J\times K}$ and respondent-level subdimension scores $S_{ik}$ obtained by weighted aggregation of harmonized responses.

\subsection{Taxonomy induction}
Inspired by TaxoAdapt \citep{kargupta2025taxoadapt}, we use an LLM to propose a data-driven taxonomy from the set of item stems. TaxoAdapt develops an iterative, LLM-based procedure for constructing and refining a multidimensional taxonomy that is aligned to the topical distribution of a target corpus (via repeated classification, expansion, and consolidation). In our setting, the ``corpus'' is the finite collection of survey question stems (and, where relevant, response-option text).

\paragraph{Step 0: researcher-defined coarse dimensions (anchors).}
Before invoking the LLM, the researcher supplies a small set of economically interpretable, coarse dimensions e.g., beliefs and constraints. These anchors serve two purposes: they constrain the space of admissible mechanisms for interpretability, and they reduce the risk that the taxonomy drifts into idiosyncratic semantic distinctions that are hard to map to economic theory.

\paragraph{Step 1: first-layer subdimension proposal within anchors.}
Conditional on the anchor dimensions and the full item inventory, the LLM proposes a first-layer set of subdimensions within each anchor. Each subdimension is accompanied by (i) a short, mechanism-oriented definition, (ii) inclusion/exclusion boundary rules, and (iii) a list of representative items that motivated the label. The LLM sees question text only (not outcomes), so taxonomy induction cannot ``fit'' behavior by construction.

\paragraph{Step 2: consolidation and audit.}
We consolidate near-duplicate candidates and impose naming and boundary consistency checks to obtain an initial taxonomy $\mathcal{T}^{(0)}$ with $K$ subdimensions. We treat the resulting taxonomy as provisional: subsequent econometric diagnostics (Section~\ref{sec:validation}) determine whether it is retained as-is or refined using the operators in Section~\ref{sec:operators}. The resulting initial taxonomy, including top-level dimensions, subdimensions, and their motivating survey items, is summarized in Appendix Table~\ref{tab:taxonomy_v1}.

\paragraph{Reproducibility.}
All prompts used for taxonomy induction and mapping are versioned and reported in Appendix~\ref{app:prompt_templates}, so the taxonomy is auditable and can be regenerated under the same protocol.

\subsection{Item-to-subdimension soft mapping on the simplex}
Given a taxonomy $\mathcal{T}$, we map each item $j$ to a sparse distribution over subdimensions. For each item stem $q_j$, the LLM outputs weights $\{w_{jk}\}_{k=1}^K$ such that
\begin{equation}
w_{jk}\ge 0
\quad\text{and}\quad
\sum_{k=1}^K w_{jk}=1
\qquad \forall j.
\label{eq:simplex}
\end{equation}
To reduce noise and enhance interpretability, we restrict mappings to be sparse (e.g., at most $m$ nonzero weights per item), dropping very small weights and renormalizing to satisfy \eqref{eq:simplex}. This soft mapping captures multi-mechanism item content while keeping item representations comparable.

\subsection{Response harmonization}
Survey items differ in response type: binary, ordinal, categorical, and numeric fields. To aggregate responses across heterogeneous items, we first map each response into a unified numeric representation $x_{ij}$ on an item-by-item basis.

Operationally, we (i) preserve ordering when meaningful (e.g., Likert scales), (ii) map categorical bins to ordered numeric codes when an ordering is substantively justified, (iii) apply monotone transforms for skewed numeric inputs when needed (e.g., \texttt{log1p}), and (iv) treat non-substantive responses (e.g., ``Prefer not to say'') as missing. For comparability across heterogeneous scales, we winsorize and standardize within the training fold and then apply the same transformation to the test fold to avoid leakage in out-of-sample evaluation. Appendix~\ref{app:harmonization} documents the full item inventory and harmonization rule used for each item.

Given $W$ and harmonized responses $\{x_{ij}\}$, we construct respondent-level subdimension scores by weighted aggregation:
\begin{equation}
S_{ik}
=
\frac{\sum_{j=1}^J w_{jk}\, x_{ij}}{\sum_{j=1}^J w_{jk}},
\label{eq:score}
\end{equation}
excluding missing $x_{ij}$ from both numerator and denominator.

\subsection{Baseline measurement system versus LLM soft mapping}
To benchmark the soft-mapping system, we also construct a baseline measurement system that assigns each item to a single dimension (hard mapping) and aggregates within dimension using standard index construction rules. The baseline serves as a parsimonious comparator.

The key difference is conceptual: the baseline enforces a one-item-one-construct restriction, whereas the LLM mapping permits sparse cross-loading on the simplex. This matters when item wording bundles mechanisms. Accordingly, we interpret LLM weights $W$ as \emph{measurement proposals} that can be validated.
\begin{itemize}[leftmargin=*]
  \item \textbf{Baseline (hard mapping).} Each item contributes to exactly one construct, yielding easy-to-interpret indices but risking construct contamination when items are multi-mechanism.
  \item \textbf{LLM (soft mapping).} Each item can load on multiple subdimensions with nonnegative weights that sum to one, making ambiguity explicit and allowing the data to reveal whether separating mechanisms improves out-of-sample validity.
\end{itemize}

\subsection{The iterative measurement loop}
\label{subsec:loop}
Sections~\ref{sec:validation}--\ref{sec:operators} turn the soft mapping in
\eqref{eq:simplex}--\eqref{eq:score} into a disciplined iteration procedure.
The LLM serves as a \emph{proposal engine}, suggesting candidate constructs and sparse item weights, while econometric diagnostics determine whether these proposals are retained, modified, or discarded based on incremental out-of-sample value and discriminant validity.

\section{Estimation and Validation Strategy}
\label{sec:validation}

This section defines the validation rules that govern any application of the measurement framework. No empirical results are reported here. Instead, we specify how candidate constructs proposed by the LLM are evaluated, refined, or discarded using econometric criteria.

We validate the measurement system along two dimensions. First, we assess whether adding a candidate subdimension improves predictive or explanatory performance out of sample relative to baseline specifications that exclude it. Second, we evaluate whether the candidate is empirically distinct, in the sense that it captures signal not already absorbed by nearby constructs. Together, these checks operationalize an econometrics-as-arbiter principle: the LLM proposes constructs and weights, while data-driven diagnostics determine which proposals are retained and which are revised.

\subsection{Incremental out-of-sample value}

For each candidate subdimension, we compute an incremental out-of-sample (OOS) gain relative to a benchmark specification without the candidate.\footnote{The protocol supports any scalar OOS metric, including $R^2$, RMSE, log loss, or AUC, depending on the downstream task. Appendix~\ref{app:ecv} provides implementation details.} We emphasize both magnitude and stability across folds. Components are classified as \emph{signal} when gains are positive and stable, as \emph{weak signal} when gains are small but stable, and as \emph{noise-like} when gains are near zero or unstable across folds.

\subsection{Discriminant-validity diagnostics}

Out-of-sample performance alone does not distinguish between true null effects and construct contamination. We therefore pair OOS triage with simple and interpretable separation diagnostics designed to detect overlap and misassignment. These include high correlations with nearby subdimensions, concentrated cross-loadings in the item-to-construct mapping, and failure of conditional contribution tests within overlap clusters. Formal definitions and recommended thresholds are provided in Appendix~\ref{app:ecv}.

\subsection{Decision rules: retain, refine, defer, or discard}
\label{subsec:decisions}

We implement a two-stage triage procedure. In Stage~1, each candidate component is labeled according to its OOS performance as signal, weak signal, or noise-like. In Stage~2, noise-like components are diagnosed as either \emph{data-limited}, reflecting insufficient coverage or support, or \emph{overlap-driven}, reflecting empirical inseparability from nearby constructs. Actions follow directly:
\begin{itemize}[leftmargin=*]
  \item \textbf{Retain} if OOS gains are stable and the component is empirically distinct, or if it passes conditional contribution tests within an overlap cluster.
  \item \textbf{Refine} if the component is noise-like and overlap-driven, using the refinement operators described below.
  \item \textbf{Defer (report as data-limited)} if the component is noise-like due to insufficient measurement coverage rather than taxonomy failure.
  \item \textbf{Discard} if the component repeatedly fails OOS triage and is not improved by targeted refinement.
\end{itemize}

\subsection{Iterative refinement operators}
\label{sec:operators}

This subsection defines the refinement operators used when a component is noise-like in incremental OOS tests \emph{and} diagnostics indicate overlap-driven contamination (e.g., persistent high correlations, concentrated cross-loadings, or failure of conditional contribution within an overlap cluster). Refinement is applied locally: we modify only the affected neighborhood of constructs and items, while holding unrelated dimensions fixed. All updates preserve the simplex constraint in \eqref{eq:simplex}.

\paragraph{Operator 1: Anchoring.}
Anchoring fixes a subset of dimensions whose measurement is treated as stable. For anchored dimensions, item-to-dimension weights are held invariant across iterations:
\[
w^{(a)}_{j d} \in \{0,1\},
\quad
\sum_{d \in \mathcal{D}_a} w^{(a)}_{j d} \le 1.
\]
Anchors serve as reference standards and prevent global relabeling during refinement.

\paragraph{Operator 2: Local splitting.}
When overlap concentrates within a parent subdimension $d$, we introduce a small set of child subdimensions $\{d_1,\ldots,d_K\}$ nested within $d$ and reallocate content locally. Re-estimation is restricted to items in the affected neighborhood $\mathcal{I}(d)$:
\[
w_{j d_k} \ge 0,\quad
\sum_{k=1}^K w_{j d_k} \le 1,
\quad \text{for } j \in \mathcal{I}(d),
\]
while anchored and unrelated dimensions remain fixed.

\paragraph{Operator 3: Constraint tightening.}
When overlap is driven by diffuse cross-loading or many small weights, we tighten mapping constraints while preserving the simplex structure. Examples include enforcing a one-primary-plus-at-most-one-secondary loading rule, thresholding small weights (followed by renormalization), and constrained relabeling for a small set of items identified as overlap drivers.

\paragraph{Stopping rule.}
Iteration stops when overlap diagnostics no longer indicate concentrated cross-loadings among retained constructs and incremental OOS gains plateau across successive refinement rounds.

\section{Data and Application Setting}
\label{sec:application}

This section describes the survey data and application used to illustrate the measurement framework. The application draws on a large-scale survey of U.S.\ public employees designed to elicit how workers value continued accruals under their incumbent defined benefit (DB) pension plans relative to a hypothetical switch to a defined contribution (DC) alternative.

\subsection{Survey design and valuation task}

Our empirical application uses the public-employee retirement plan survey studied by \citet{giesecke2022much}. The survey is designed to measure how public employees value future DB accruals relative to a forward-looking switch to a DC plan on a hard-freeze basis. Under this scenario, already-accrued DB benefits are preserved, but no additional DB benefits accrue going forward. In a DB plan, the employer guarantees retirement benefit payments and bears investment risk, whereas in a DC plan the employer makes fixed contributions into an individual account and employees choose how to invest.

The valuation task therefore maps into a compensating differential. Respondents are asked whether they would enroll in a hypothetical DC plan that replaces future DB accruals and, conditional on enrollment, what minimum employer contribution rate, expressed as a percentage of payroll, would make them indifferent between remaining in the DB plan and switching to the DC alternative \citep{giesecke2022much}. These responses provide direct measures of both discrete acceptance and the intensity of valuation.

\subsection{Survey administration and sampling frame}

\citet{giesecke2022much} field the survey via email to public employees in U.S.\ state and local government, state higher education, and K--12 school districts across 16 states. They compile approximately 396{,}948 publicly available email addresses and obtain 7{,}674 completed responses. After accounting for bounce-backs from inactive or inaccurate addresses, this corresponds to an adjusted response rate of approximately 2.1\%. While modest, the sample size is large for a detailed pension valuation survey and supports rich heterogeneity analysis.

\subsection{Instrument structure and measured constructs}

Beyond the DB--DC indifference elicitation, the survey collects detailed information on employment and pension status, perceptions and beliefs about the incumbent plan and the DC alternative, financial literacy, and a broad set of demographic and socioeconomic characteristics. Perception and belief measures include assessments of plan generosity and perceived financial stability, which are central to the valuation of retirement benefits.

Financial literacy is measured using a short battery of basic household finance questions. \citet{giesecke2022much} construct a financial literacy score defined as the percentage of questions answered correctly, ranging from 0 to 100, and refer to this block as questions Q34--Q40 in the survey instrument. In addition to subjective perceptions, the analysis links respondents to objective measures of plan generosity. Specifically, employer service cost as a percentage of payroll is obtained from pension plan financial disclosures and used as a proxy for the actuarial value of DB accruals.

\subsection{Main outcomes and baseline covariates}

The application focuses on two outcomes that correspond directly to the survey's core elicitation. The first is acceptance of the hypothetical DC option. The second is the minimum required employer contribution rate, expressed as a percentage of payroll, measured among respondents who indicate willingness to switch. The paper documents high acceptance rates overall and summarizes the distribution of required contribution rates among accepters.

In downstream specifications, we follow \citet{giesecke2022much} in using a comprehensive set of baseline covariates. These include demographics and job characteristics, such as age and years of service; socioeconomic controls, including household income; cognitive proxies, such as financial literacy and educational attainment; and plan-level measures, including employer service cost as a percentage of payroll. The richest specifications include state fixed effects.


\section{Empirical Results}
\label{sec:workflow}

This section reports the empirical results from applying the AI-assisted measurement framework to the public-employee retirement plan survey. The workflow links (i) LLM-based soft mapping of survey items into a multidimensional measurement taxonomy, (ii) harmonization and construction of respondent-level measurement scores, and (iii) out-of-sample (OOS) econometric tests that distinguish predictive \emph{signal} from \emph{noise}. OOS diagnostics serve both as evaluation tools and as discipline mechanisms for iteration: the taxonomy is refined only when diagnostics indicate systematic construct overlap that can be resolved through a more interpretable decomposition. Supporting mapping artifacts and refinement allocations are reported in the Appendix.

\subsection{Out-of-sample incremental validity of measurement components}
\label{sec:workflow_oos}

We evaluate each constructed subdimension using incremental out-of-sample validity tests (Section~\ref{sec:validation}). For the binary acceptance outcome, we report $\Delta$AUC and $\Delta$LogLoss. For the required employer contribution rate, estimated on the acceptor subsample, we report $\Delta R^2$ and $\Delta$RMSE (RMSE is reported as baseline minus augmented). Components are interpreted jointly by magnitude and stability across folds.

Table~\ref{tab:oos_accept_full} reports full cross-validated incremental validity results for acceptance. Two components stand out. \textit{service\ tenure\ lockin} delivers the largest and most stable gains ($\Delta$AUC $\approx 0.114$), and \textit{financial\ literacy} also contributes strongly ($\Delta$AUC $\approx 0.082$). By contrast, \textit{employment\ context} and the initial \textit{perceived\_generosity} aggregate behave as noise-like.

\begin{table}[!htbp]
\centering
\caption{Out-of-sample incremental validity for acceptance}
\label{tab:oos_accept_full}
\small
\setlength{\tabcolsep}{6pt}
\renewcommand{\arraystretch}{1.15}

\begin{threeparttable}
\begin{tabular}{lccc cc cc c}
\toprule
& \multicolumn{2}{c}{Coverage} 
& \multicolumn{2}{c}{AUC} 
& \multicolumn{2}{c}{Log Loss} 
& Stability \\
\cmidrule(lr){2-3}
\cmidrule(lr){4-5}
\cmidrule(lr){6-7}
\cmidrule(lr){8-8}

Subdimension 
& Items 
& $N$ 
& $\Delta$ Mean 
& SD 
& $\Delta$ Mean 
& SD 
& Share Improve \\

\midrule
Service tenure lock-in 
& 2 & 5,524 
& 0.114 & 0.030 
& $-0.0061$ & 0.0022 
& 1.00 \\

Financial literacy 
& 5 & 5,285 
& 0.082 & 0.024 
& $-0.0030$ & 0.0020 
& 0.92 \\

Income / wealth buffer 
& 4 & 5,438 
& 0.036 & 0.040 
& $-0.0004$ & 0.0011 
& 0.60 \\

Retirement horizon 
& 3 & 5,521 
& 0.024 & 0.031 
& $-0.0002$ & 0.0007 
& 0.76 \\

Demographics 
& 4 & 5,524 
& 0.019 & 0.032 
& 0.0001 & 0.0013 
& 0.72 \\

Employment context 
& 3 & 5,524 
& $-0.002$ & 0.028 
& 0.0004 & 0.0008 
& 0.32 \\

Perceived generosity 
& 3 & 4,045 
& $-0.003$ & 0.004 
& 0.0001 & 0.0003 
& 0.40 \\

\bottomrule
\end{tabular}

\begin{tablenotes}[flushleft]\footnotesize
\item \emph{Notes:} Entries report cross-validated out-of-sample improvements relative to the
baseline model that includes standard covariates but excludes the listed subdimension.
$\Delta$AUC and $\Delta$Log Loss are averaged across 25 folds.``Share Improve'' is the fraction of folds with positive improvement.
\end{tablenotes}
\end{threeparttable}
\end{table}

Table~\ref{tab:oos_rate_full} reports the corresponding results for the required contribution rate among accepters. The dominance of \textit{service\ tenure\_lockin} persists on this intensive margin ($\Delta R^2_{\text{OOS}} \approx 0.0089$), while most other components contribute weakly or not at all.

\begin{table}[!htbp]
\centering
\caption{Out-of-sample incremental validity for required contribution rate (accepters only)}
\label{tab:oos_rate_full}
\small
\setlength{\tabcolsep}{6pt}
\renewcommand{\arraystretch}{1.15}

\begin{threeparttable}
\begin{tabular}{lccc cc cc c}
\toprule
& \multicolumn{2}{c}{Coverage} 
& \multicolumn{2}{c}{$R^2$} 
& \multicolumn{2}{c}{RMSE} 
& Stability \\
\cmidrule(lr){2-3}
\cmidrule(lr){4-5}
\cmidrule(lr){6-7}
\cmidrule(lr){8-8}

Subdimension 
& Items 
& $N$ 
& $\Delta$ Mean 
& SD 
& $\Delta$ Mean 
& SD 
& Share Improve \\

\midrule
Service tenure lock-in 
& 2 & 5,524 
& 0.0089 & 0.0044 
& $-0.000896$ & 0.000446 
& 0.96 \\

Financial literacy 
& 5 & 5,285 
& 0.0009 & 0.0026 
& $-0.000095$ & 0.000257 
& 0.76 \\

Retirement horizon 
& 3 & 5,521 
& 0.0003 & 0.0028 
& $-0.000025$ & 0.000285 
& 0.68 \\

Employment context 
& 3 & 5,524 
& 0.0006 & 0.0021 
& $-0.000058$ & 0.000206 
& 0.64 \\

Income / wealth buffer 
& 4 & 5,438 
& $-0.0004$ & 0.0012 
& 0.000036 & 0.000116 
& 0.64 \\

Perceived generosity 
& 3 & 4,045 
& $-0.0005$ & 0.0010 
& 0.000049 & 0.000099 
& 0.52 \\

Demographics 
& 4 & 5,524 
& $-0.0004$ & 0.0008 
& 0.000039 & 0.000086 
& 0.48 \\

\bottomrule
\end{tabular}

\begin{tablenotes}[flushleft]\footnotesize
\item \emph{Notes:} Entries report cross-validated out-of-sample improvements relative to the baseline model for the required contribution rate outcome, estimated on the acceptor subsample.
$\Delta R^2$ and $\Delta$RMSE are averaged across 25 folds.
``Share Improve'' is the fraction of folds with positive improvement in the corresponding metric.
RMSE is reported as baseline minus augmented (so positive values indicate improvement).
\end{tablenotes}
\end{threeparttable}
\end{table}

\subsection{Overlap diagnostics and construct contamination}
\label{sec:workflow_overlap}

Incremental OOS performance does not by itself distinguish between true null effects and construct contamination. We therefore complement OOS triage with overlap diagnostics that proxy for discriminant validity.

Table~\ref{tab:overlap_pairs} reports all constructed subdimension pairs with $|\rho|\ge 0.85$. A single, pronounced overlap emerges: \textit{financial\_literacy} and \textit{perceived\_generosity} have correlation 0.948. This near-collinearity is consistent with the cross-loading pattern observed in the item-level mapping and motivates targeted refinement rather than wholesale discarding.

\begin{table}[!htbp]
\centering
\caption{High correlations among constructed subdimension scores}
\label{tab:overlap_pairs}
\small
\setlength{\tabcolsep}{8pt}
\renewcommand{\arraystretch}{1.15}

\begin{threeparttable}
\begin{tabular}{llc}
\toprule
Subdimension A & Subdimension B & Correlation \\ 
\midrule
Financial literacy & Perceived generosity & 0.948 \\
\bottomrule
\end{tabular}

\begin{tablenotes}[flushleft]\footnotesize
\item \emph{Notes:} The table reports pairs of constructed subdimension scores with absolute correlation
$|\rho| \ge 0.85$, computed using respondent-level scores.
High correlations are interpreted as evidence of potential construct overlap and motivate the targeted
refinement procedures described in Section~\ref{sec:workflow_refine}.
\end{tablenotes}
\end{threeparttable}
\end{table}

\subsection{Targeted refinement and re-evaluation}
\label{sec:workflow_refine}

Guided by the overlap diagnostics, we implement a targeted refinement that anchors the canonical Financial Literacy battery (Q34--Q40) and decomposes \texttt{perceived\_generosity} into more granular belief channels. The allocation of item weight mass across child constructs is reported in Appendix Table~\ref{app:tab:generosity_split_allocation}, and the refined OOS incremental validity results are reported in Appendix Table~\ref{app:tab:oos_v2_generosity_split}. The refinement prompt and constraints used to implement the split are documented in Appendix Figure~\ref{app:fig:prompt_refinement}.

\subsection{Robustness and placebo evidence}
\label{sec:workflow_robust}

We assess robustness along three dimensions: sparsification thresholds, alternative score construction rules, and permutation placebo tests. Robustness tables are reported in Appendix~\ref{app:robustness_suite}. In particular, Appendix Table~\ref{tab:robust_tau_topm} shows that overlap diagnostics and incremental OOS performance are stable across sparsification thresholds and top-$m$ constraints, while Appendix Table~\ref{tab:robust_placebo} reports permutation placebo results that destroy semantic alignment while preserving mechanical complexity.

\begin{figure}[!htbp]\centering
\begin{subfigure}{0.48\linewidth}\centering
\includegraphics[width=\linewidth]{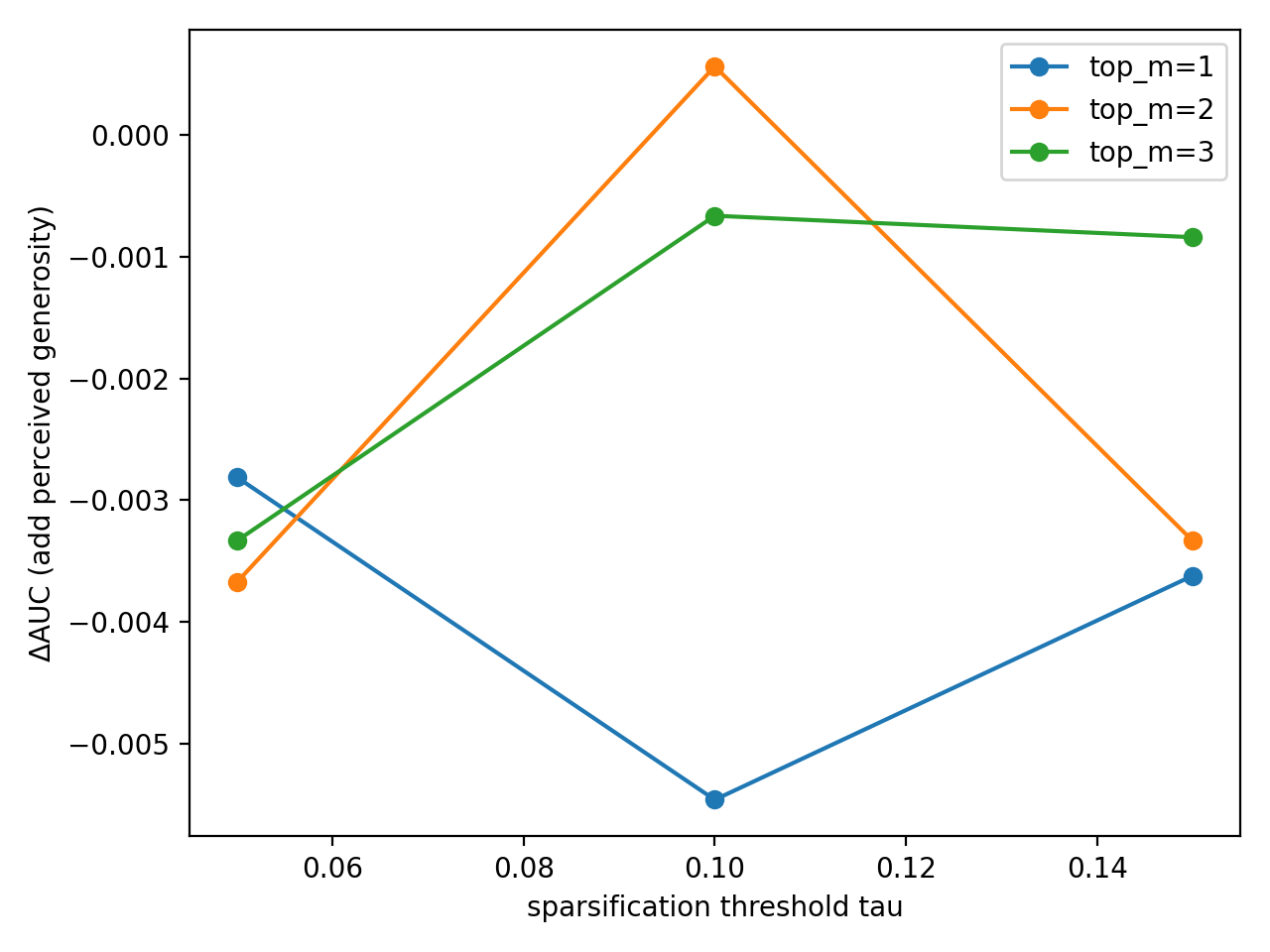}
\caption{Acceptance: $\Delta$AUC across $\tau$ and top-$m$.}
\end{subfigure}\hfill
\begin{subfigure}{0.48\linewidth}\centering
\includegraphics[width=\linewidth]{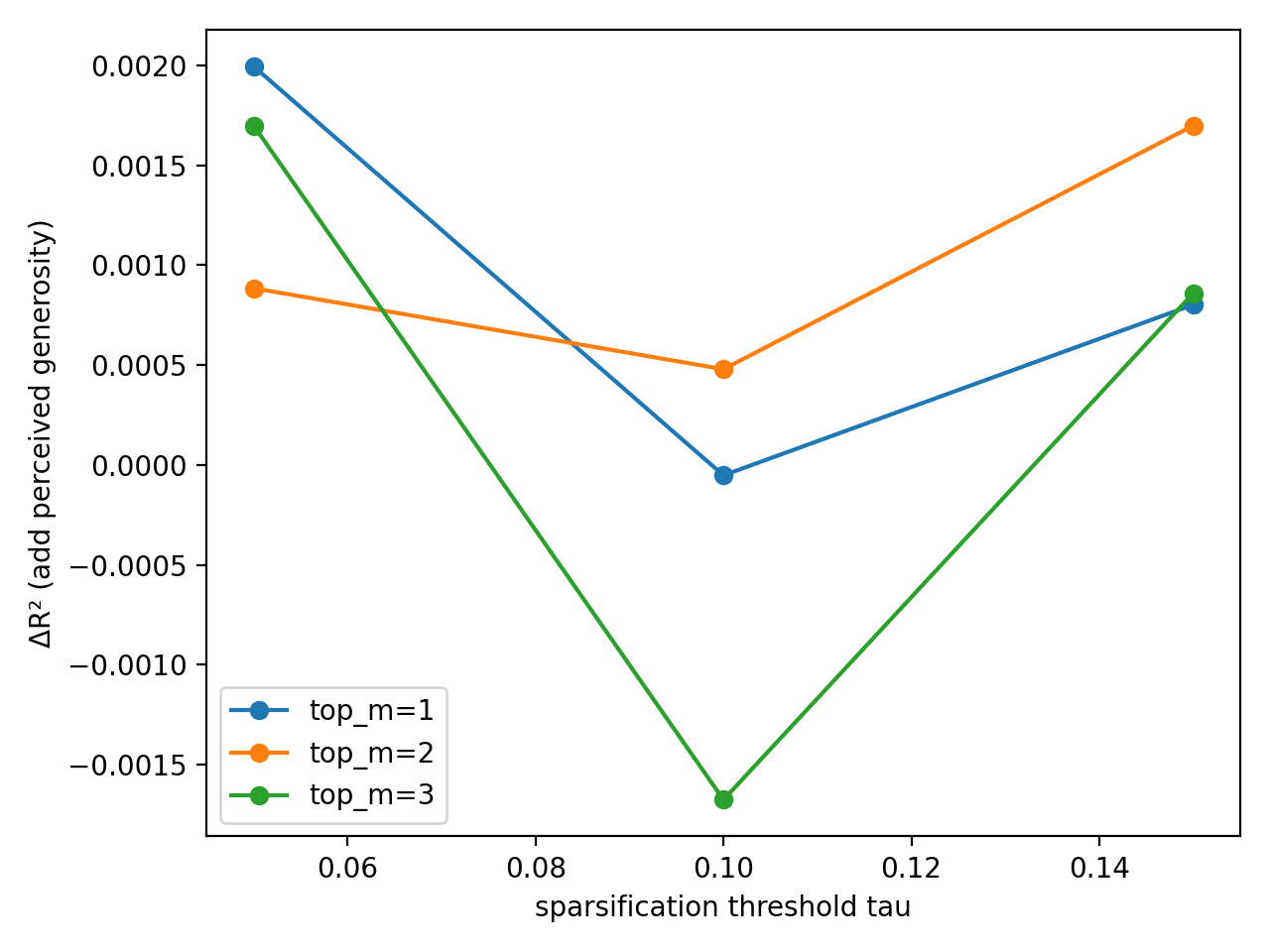}
\caption{Rate (accepters): $\Delta R^2$ across $\tau$ and top-$m$.}
\end{subfigure}
\caption{Threshold robustness for the initial soft mapping.}
\label{fig:robust_tau_topm}
\end{figure}

\begin{figure}[!htbp]\centering
\begin{subfigure}{0.48\linewidth}\centering
\includegraphics[width=\linewidth]{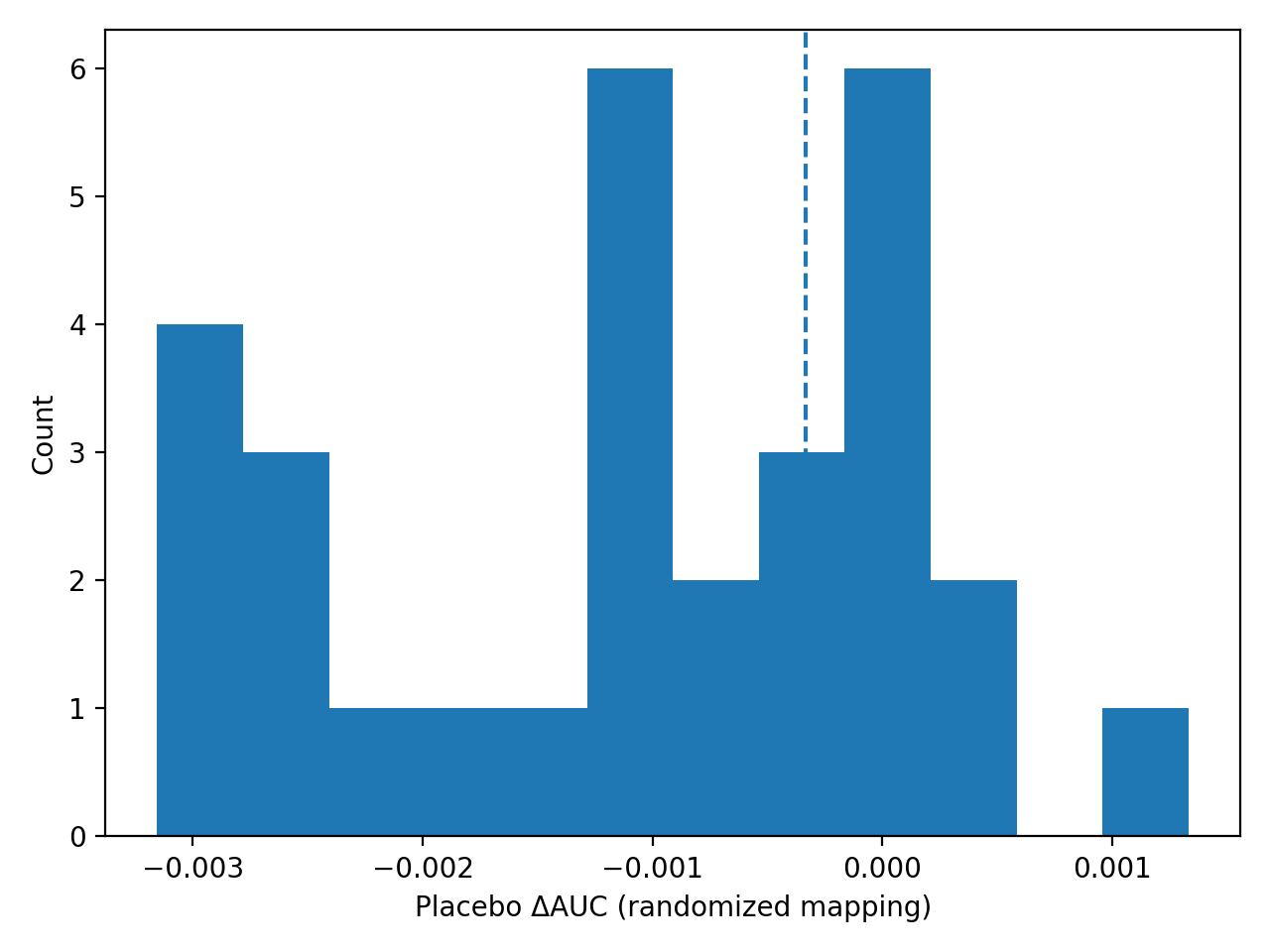}
\caption{Placebo distribution for $\Delta$AUC.}
\end{subfigure}\hfill
\begin{subfigure}{0.48\linewidth}\centering
\includegraphics[width=\linewidth]{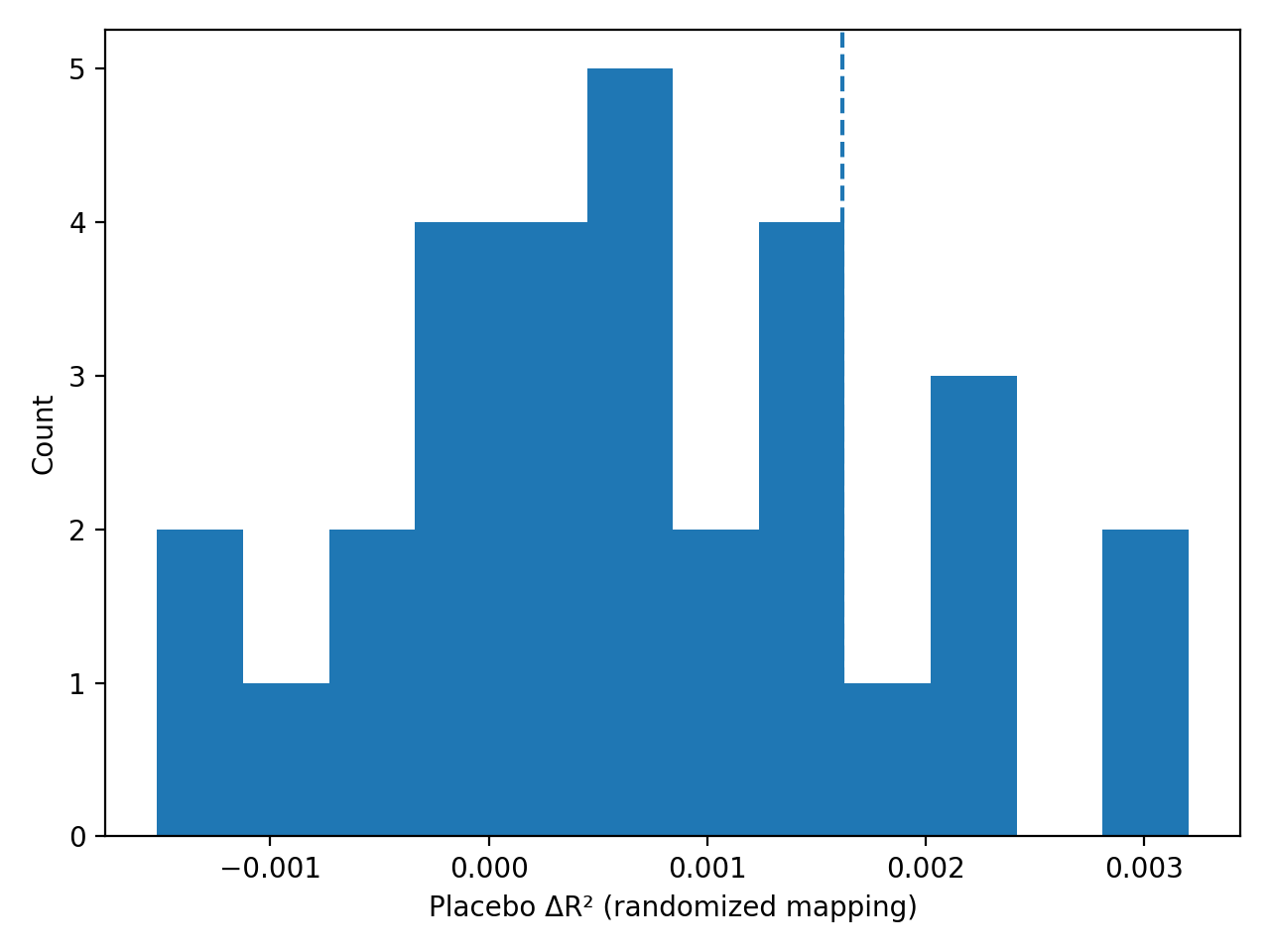}
\caption{Placebo distribution for $\Delta R^2$.}
\end{subfigure}
\caption{Permutation placebo: randomized mapping versus observed mapping. The dashed vertical line marks the observed mapping.}
\label{fig:placebo_hists}
\end{figure}

Figure~\ref{fig:robust_tau_topm} summarizes threshold robustness across weight thresholds $\tau$ and top-$m$ constraints. Figure~\ref{fig:placebo_hists} reports placebo distributions that preserve mechanical complexity while destroying semantic alignment. Robustness tables, including Table~\ref{tab:robust_scoring_rules} and other grid summaries, are available in Appendix~\ref{app:robustness_suite}.

For completeness, Appendix Table~\ref{app:tab:triage-summary} consolidates the full out-of-sample triage across all constructed subdimensions, reporting incremental performance, stability classifications, and refinement outcomes. This table summarizes how the diagnostic rules map candidate constructs into retained signal, weak signal, and discarded noise.

\section{Discussion}
\label{sec:discussion}

The empirical results in Section~\ref{sec:workflow} illustrate how the framework operates in practice and how it can be used to extract, validate, and refine latent economic mechanisms encoded in survey wording. The application demonstrates how a rich instrument can be converted into a transparent and testable measurement system.

Several patterns emerge consistently from the results. First, economically meaningful signal is highly concentrated. Across both outcomes studied---acceptance of a hypothetical DC option and the minimum required employer contribution rate---the strongest and most stable incremental information is captured by the \texttt{service\_tenure\_lockin} channel. This dimension survives out-of-sample (OOS) validation across folds and remains robust under refinement, indicating that career-stage incentives and lock-in considerations represent the most reliably measured mechanism in the instrument. Second, financial literacy content contributes meaningfully to predicting acceptance but adds little stable incremental value for the required contribution rate among accepters. Third, several intuitively appealing constructs behave as noise-like when evaluated as aggregate indices, but targeted refinement can recover interpretable weak-signal belief channels, including a plan-stability channel that delivers materially larger incremental gains for acceptance.

These findings align naturally with, and build upon, the empirical patterns documented in \citet{giesecke2022much}. Their analysis shows strong heterogeneity by years of service and documents that financial literacy is associated with acceptance but largely unrelated to the intensive margin of valuation. Our framework preserves these baseline relationships for comparability, but adds a distinct measurement layer that formalizes them. Rather than relying on pre-specified, hard-coded indices, we allow survey items to load flexibly on multiple latent subdimensions and then use OOS diagnostics to determine which of these dimensions are empirically supported. In this sense, the framework complements existing applied work by converting reduced-form patterns into explicit, validated measurement objects, while avoiding over-interpretation of dimensions that do not survive validation.

From a practical perspective, the framework offers applied researchers a way to work systematically with complex survey instruments. Because the mapping is defined at the level of item text rather than estimation samples, it is portable across survey waves and related instruments. The diagnostics identify which parts of an instrument contain reliable signal for particular outcomes and which domains are too coarse, redundant, or weakly connected to behavior. In the pension setting, the results highlight the central role of lock-in mechanisms for both extensive and intensive margins, while suggesting that informational content is most relevant for take-up decisions rather than valuation. More generally, the approach can guide survey redesign by indicating where additional items are likely to improve measurement and where existing batteries may be redundant.

Several limitations are worth noting. The framework evaluates measurement quality relative to specific outcomes, so constructs that appear noise-like here may be relevant in other applications. Finite instruments constrain the granularity of recoverable mechanisms, and OOS gains are necessarily modest in weak-signal settings. Moreover, while semantic tools reduce the burden of parsing item content, economic judgment remains essential in selecting anchors, interpreting refined constructs, and assessing policy relevance. These limitations are inherent to measurement rather than specific to the use of language models.

Despite these caveats, the results demonstrate that combining semantic analysis with econometric validation can sharpen economic measurement in a transparent and replicable way. By making ambiguity explicit, disciplining flexibility with OOS diagnostics, and anchoring interpretation in established economic theory, the framework provides a general approach for extracting mechanism-relevant information from survey instruments while remaining firmly grounded in empirical economics.

\section{Conclusion}
\label{sec:conclusion}

We propose a diagnostic-driven, AI-assisted framework for economic measurement that starts from survey instruments and ends with validated, portable measurement objects. The framework combines simplex-constrained soft mapping from item text to latent subdimensions, unified response harmonization across heterogeneous item formats, and an econometric validation protocol based on out-of-sample incremental performance and overlap diagnostics. Crucially, outputs from large language models are treated as measurement proposals rather than as ground truth, with transparent decision rules determining which constructs are retained, refined, or rejected as noise. Applied to a public-employee retirement plan survey, the framework shows that only a small subset of candidate subdimensions delivers stable incremental information beyond baseline measurement: lock-in mechanisms generate the strongest gains for both acceptance and required contribution rates, while financial literacy contributes meaningfully to acceptance but only weakly to the intensive margin. These patterns align with existing qualitative interpretations in the application setting while adding a validation-based interpretation, showing that weak intensive-margin relationships persist under cross-validation rather than reflecting in-sample specification choices. More broadly, the contribution is methodological. The framework provides a practical way to translate high-dimensional survey instruments into economic measurements that are auditable, reusable, and disciplined by standard econometric criteria rather than by semantic richness or in-sample fit. Future work can extend the validation suite to other outcomes, settings, and survey designs, and examine when soft mappings generalize across instruments that differ in wording, ordering, or response formats.

\medskip
\noindent \textbf{Disclosure.}
None of the authors have financial or other ties related to the research project.

\bibliographystyle{apalike}
\bibliography{references}

\newpage
\appendix
\section*{Appendix}
\addcontentsline{toc}{section}{Appendix}
\renewcommand{\thetable}{A\arabic{table}}
\setcounter{table}{0}
\setcounter{figure}{0}

\section{Prompt templates and reproducibility}
\label{app:taxonomy}

\subsection{Prompt templates}
\label{app:prompt_templates}
This appendix records the prompt templates used to (i) induce a first-layer taxonomy of subdimensions from survey item stems and (ii) produce a constrained item-to-subdimension mapping. The goal is to make the LLM component \emph{auditable} and \emph{replicable}: a reader should be able to rerun the same protocol on the same instrument and recover the same intermediate artifacts up to stochastic variation.

\subsubsection{A. Taxonomy induction prompt}
\noindent\textbf{Researcher inputs.} The researcher provides:
\begin{itemize}[leftmargin=*]
  \item A small set of coarse anchor dimensions (first-layer buckets), each with a one-sentence economic interpretation.
  \item The full list of survey item stems (and, where relevant, response options).
  \item A requested granularity (target number of subdimensions per anchor, or a maximum number).
\end{itemize}

\noindent\textbf{Prompt template (system/user content).}
\begin{verbatim}
You are helping construct an interpretable measurement taxonomy for an economic survey.

INPUTS:
1) Coarse anchor dimensions (fixed by the researcher):
- [ANCHOR_1]: [one-sentence definition]
- [ANCHOR_2]: [one-sentence definition]
...

2) Survey items (question stems, text only; do NOT use outcomes):
- Item 1: [text]
- Item 2: [text]
...

TASK:
For each anchor dimension, propose first-layer SUBDIMENSIONS that are (i) mechanism-oriented,
(ii) mutually as distinct as possible, and (iii) likely to be useful for economic interpretation.

For each proposed subdimension, output:
- Name (snake_case)
- Anchor dimension it belongs to
- Definition (2-3 sentences)
- Inclusion rules (bullet points)
- Exclusion rules / boundary with closest alternative (bullet points)
- Representative items (list item IDs)

CONSTRAINTS:
- Use only the item text (no outcomes, no response frequencies).
- Prefer fewer, clearer subdimensions to many narrow ones.
- Avoid duplicates and jingle-jangle labels.
- If an item plausibly spans multiple subdimensions, note this explicitly.

OUTPUT FORMAT:
Return a JSON object with anchors as keys and a list of subdimensions as values.

a) taxonomy: { ... }
 b) brief rationale: 5-10 bullet points on the design choices.

a) and b) only.
\end{verbatim}

\subsubsection{B. Soft-mapping prompt}
\noindent\textbf{Prompt template.}
\begin{verbatim}
You are given a fixed taxonomy of constructs and a list of survey item stems.
Your task is to assign each item a sparse simplex-weighted mixture over constructs.

INPUTS:
1) Taxonomy (fixed):
- [CONSTRUCT_1]: [definition + boundary rules]
- [CONSTRUCT_2]: [definition + boundary rules]
...

2) Item stem:
- Item [ID]: [text]

TASK:
Output weights w_k such that:
- w_k >= 0 for all k
- sum_k w_k = 1
- sparsity: at most [m] nonzero weights

Also provide:
- One-sentence rationale
- One-sentence "closest alternative and why not"

OUTPUT FORMAT (one item at a time):
- weights: {construct_name: weight, ...}
- rationale: ...
- not_this: ...
\end{verbatim}

\subsubsection{C. Researcher post-processing and checks}
After obtaining raw LLM outputs, we apply deterministic checks before using them downstream:
\begin{itemize}[leftmargin=*]
  \item \textbf{Deduplication:} merge near-duplicate subdimension labels with overlapping definitions.
  \item \textbf{Boundary consistency:} ensure each subdimension has a clear ``not-this'' neighbor.
  \item \textbf{Sparsification:} drop very small item weights and renormalize to satisfy the simplex.
  \item \textbf{Versioning:} archive (prompt, model, temperature, taxonomy version) for each iteration.
\end{itemize}

\subsection{Initial taxonomy summary}
Table~\ref{tab:taxonomy_v1} summarizes the initial taxonomy produced by the initial soft mapping pass, including the top-level dimension, subdimension name, and the list of harmonized items contributing to each construct score.

\footnotesize

\begin{longtable}{p{2.2cm} p{2.6cm} p{0.9cm} p{4.8cm} p{6.0cm}}
\caption{Initial taxonomy used in the initial LLM soft-mapping pass. Items are survey stems mapped to each construct (after harmonization).}\label{tab:taxonomy_v1}\\ 
\toprule
Top-level & Subdimension & \# & Items & Definition\\
\midrule
\endfirsthead
\toprule
Top-level & Subdimension & \# & Items & Definition\\
\midrule
\endhead
\midrule \multicolumn{5}{r}{\emph{Continued on next page}}\\ 
\endfoot
\bottomrule
\endlastfoot
Cognition\_time & financial\_literacy & 5 & tilde\_Q14, tilde\_Q16, tilde\_Q17, tilde\_Q18, tilde\_Q33 & Basic financial knowledge and numeracy for evaluating pension/portfolio tradeoffs.\\
Controls & demographics & 4 & tilde\_Q42, tilde\_Q44, tilde\_Q7, tilde\_Q8 & Age, gender, education and related background characteristics.\\
Econ\_constraints & income\_wealth\_buffer & 4 & tilde\_Q17, tilde\_Q19, tilde\_Q6, tilde\_Q7 & Liquidity, savings buffers, and household resources affecting willingness to bear risk.\\
Controls & employment\_context & 3 & tilde\_Q4, tilde\_Q5, tilde\_Q6 & Work intensity and employment setting that affect plan participation/constraints.\\
Econ\_constraints & retirement\_horizon & 3 & tilde\_Q15, tilde\_Q19, tilde\_Q9 & Proximity to retirement and planning horizon that shape valuation of future accruals.\\
DB\_beliefs & perceived\_generosity & 3 & tilde\_Q14, tilde\_Q16, tilde\_Q18 & Beliefs about how generous current DB plan benefits are relative to alternatives.\\
Econ\_constraints & service\_tenure\_lockin & 2 & tilde\_Q15, tilde\_Q4 & Job tenure and pension lock-in (years of service, accrued benefits, switching frictions).\\
Cognition\_time & discounting & 1 & tilde\_Q9 & Time preference / patience affecting intertemporal tradeoffs.\\
Econ\_constraints & health\_risk & 1 & tilde\_Q10 & Health risk and expected healthcare needs affecting retirement preferences.\\
DB\_beliefs & perceived\_stability & 1 & tilde\_Q33 & Beliefs about DB plan solvency and benefit security.\\
\end{longtable}
\normalsize

\subsection{Item-level mapping: baseline vs.\ LLM soft mapping}

\begingroup
\scriptsize
\setlength{\tabcolsep}{7pt}
\renewcommand{\arraystretch}{1.30}

\begin{longtable}{p{0.7cm} p{2.7cm} p{7.1cm} p{1.6cm}}
\caption{Baseline hard mapping and the LLM soft mapping weights for each survey item stem.}\label{tab:mapping_codebook}\\ 
\toprule
Item & Baseline & LLM soft mapping & Use\\
\midrule
\endfirsthead
\toprule
Item & Baseline  & LLM soft mapping & Use\\
\midrule
\endhead
\midrule \multicolumn{4}{r}{\emph{Continued on next page}}\\ 
\endfoot
\bottomrule
\endlastfoot
Q1 & CONTROLS & CONTROLS.employment\_context:1.00 & Control\\
Q2 & CONTROLS & CONTROLS.employment\_context:1.00 & Control\\
Q3 & CONTROLS & CONTROLS.employment\_context:1.00 & Control\\
Q4 & ECON\_CONSTRAINTS & ECON\_CONSTRAINTS.service\_tenure\_lockin:0.90; CONTROLS.employment\_context:0.10 & Mechanism\\
Q5 & CONTROLS & CONTROLS.employment\_context:1.00 & Control\\
Q6 & ECON\_CONSTRAINTS & ECON\_CONSTRAINTS.income\_wealth\_buffer:0.95; CONTROLS.employment\_context:0.05 & Mechanism\\
Q7 & ECON\_CONSTRAINTS & ECON\_CONSTRAINTS.income\_wealth\_buffer:0.95; CONTROLS.demographics:0.05 & Mechanism\\
Q8 & CONTROLS & CONTROLS.demographics:1.00 & Control\\
Q9 & ECON\_CONSTRAINTS & ECON\_CONSTRAINTS.retirement\_horizon:0.90; COGNITION\_TIME.discounting:0.10 & Mechanism\\
Q10 & ECON\_CONSTRAINTS & ECON\_CONSTRAINTS.health\_risk:1.00 & Mechanism\\
Q11 & CONTROLS & CONTROLS.plan\_enrollment\_type:1.00 & Control\\
Q12 & CONTROLS & CONTROLS.plan\_identifier:1.00 & Control\\
Q13 & CONTROLS & CONTROLS.plan\_identifier:1.00 & Control\\
Q14 & DB\_BELIEFS & DB\_BELIEFS.perceived\_generosity:0.75; COGNITION\_TIME.financial\_literacy:0.25 & Mechanism\\
Q15 & ECON\_CONSTRAINTS & ECON\_CONSTRAINTS.retirement\_horizon:0.65; ECON\_CONSTRAINTS.service\_tenure\_lockin:0.35 & Mechanism\\
Q16 & DB\_BELIEFS & DB\_BELIEFS.perceived\_generosity:0.75; COGNITION\_TIME.financial\_literacy:0.25 & Mechanism\\
Q17 & ECON\_CONSTRAINTS & ECON\_CONSTRAINTS.income\_wealth\_buffer:0.90; COGNITION\_TIME.financial\_literacy:0.10 & Mechanism\\
Q18 & DB\_BELIEFS & DB\_BELIEFS.perceived\_generosity:0.60; COGNITION\_TIME.financial\_literacy:0.40 & Mechanism\\
Q19 & ECON\_CONSTRAINTS & ECON\_CONSTRAINTS.income\_wealth\_buffer:0.85; ECON\_CONSTRAINTS.retirement\_horizon:0.15 & Mechanism\\
Q20 & OUTCOMES & OUTCOMES.switch\_accept:1.00 & Outcome\\
Q21 & OUTCOMES & OUTCOMES.switch\_accept:1.00 & Outcome\\
Q22 & OUTCOMES & OUTCOMES.switch\_accept:1.00 & Outcome\\
Q23 & OUTCOMES & OUTCOMES.switch\_accept:1.00 & Outcome\\
Q24 & OUTCOMES & OUTCOMES.switch\_accept:1.00 & Outcome\\
Q25 & OUTCOMES & OUTCOMES.switch\_accept:1.00 & Outcome\\
Q26 & OUTCOMES & OUTCOMES.switch\_accept:1.00 & Outcome\\
Q27 & OUTCOMES & OUTCOMES.switch\_accept:1.00 & Outcome\\
Q28 & OUTCOMES & OUTCOMES.switch\_accept:1.00 & Outcome\\
Q29 & OUTCOMES & OUTCOMES.switch\_accept:1.00 & Outcome\\
Q30 & OUTCOMES & OUTCOMES.switch\_accept:1.00 & Outcome\\
Q31 & OUTCOMES & OUTCOMES.switch\_threshold:1.00 & Outcome\\
Q32 & ECON\_CONSTRAINTS & ECON\_CONSTRAINTS.health\_risk:0.70; ECON\_CONSTRAINTS.income\_wealth\_buffer:0.30 & Mechanism\\
Q33 & DB\_BELIEFS & DB\_BELIEFS.perceived\_stability:0.95; COGNITION\_TIME.financial\_literacy:0.05 & Mechanism\\
Q34 & COGNITION\_TIME & COGNITION\_TIME.financial\_literacy:1.00 & Mechanism\\
Q35 & COGNITION\_TIME & COGNITION\_TIME.financial\_literacy:1.00 & Mechanism\\
Q36 & COGNITION\_TIME & COGNITION\_TIME.financial\_literacy:1.00 & Mechanism\\
Q37 & COGNITION\_TIME & COGNITION\_TIME.financial\_literacy:1.00 & Mechanism\\
Q38 & COGNITION\_TIME & COGNITION\_TIME.financial\_literacy:1.00 & Mechanism\\
Q39 & COGNITION\_TIME & COGNITION\_TIME.discounting:0.95; COGNITION\_TIME.financial\_literacy:0.05 & Mechanism\\
Q40 & COGNITION\_TIME & COGNITION\_TIME.discounting:0.95; COGNITION\_TIME.financial\_literacy:0.05 & Mechanism\\
Q41 & CONTROLS & CONTROLS.demographics:1.00 & Control\\
Q42 & CONTROLS & CONTROLS.demographics:1.00 & Control\\
Q43 & CONTROLS & CONTROLS.demographics:1.00 & Control\\
Q44 & CONTROLS & CONTROLS.demographics:1.00 & Control\\
Q45 & CONTROLS & CONTROLS.demographics:1.00 & Control\\
Q46 & CONTROLS & CONTROLS.open\_ended\_admin:1.00 & Control\\
\end{longtable}
\endgroup
\normalsize

\subsection{Refinement rubric}
The following rubric text illustrates the mechanism-level constraints imposed in the targeted refinement step.

\begin{verbatim}
Goal: Refine LLM mapping to cleanly separate two constructs that were highly correlated in the initial pass:
(A) Cognition_time.financial_literacy  vs  (B) DB_beliefs.perceived_generosity.

Operational definitions:
A. financial_literacy (Knowledge & ability):
- Recognizing/understanding basic financial concepts (interest, inflation, risk diversification).
- Ability to compute/compare returns/fees or understand tradeoffs.
- Correct/incorrect knowledge questions count here.
NOT A: self-reported awareness of plan rules or perceptions of generosity.

B. perceived_generosity (Perceived plan generosity / employer contribution environment):
- Perceived generosity/attractiveness of employer match, benefits, or plan attractiveness.
- Perceptions/expectations about contributions or plan quality.
- Awareness of plan features counts ONLY if explicitly tied to "how generous/attractive" (otherwise route to Plan_Awareness).

Important boundary rule:
- If an item is about knowing plan rules, contribution rate, match, vesting, etc. WITHOUT a judgment of generosity, classify as Plan_Awareness (new subdimension), not A or B.

Labeling constraints (to reduce noise/collinearity):
1) For each item, output exactly ONE primary subdimension (weight 1.0).
2) Optionally output ONE secondary subdimension with weight in [0.05, 0.20] ONLY if the item clearly spans two constructs.
3) If unclear, set secondary = none.
4) Provide a one-sentence rationale and a one-sentence "why not" addressing the closest alternative label.

Output schema per item:
TopDimension.Subdimension: weight;  (optional secondary) TopDimension.Subdimension: weight
Rationale: ...
Not-this: ...
\end{verbatim}

\section{Survey Item Harmonization}
\label{app:harmonization}

Table~\ref{tab:item_harmonization} lists each survey item, its response type, and the harmonization rule used to construct the unified numeric response $x_{ij}$. Items not used in the analysis sample are marked accordingly. All winsorization and standardization steps are executed within the training fold in cross-validation and then applied to the test fold.

\footnotesize

\begin{longtable}{p{0.7cm} p{7.0cm} p{2.8cm} p{5.3cm}}
\caption{Survey item inventory and harmonization rules used to construct unified numeric responses $x_{ij}$.}\label{tab:item_harmonization}\\ 
\toprule
Item & Question (short) & Response type & Harmonization rule\\
\midrule
\endfirsthead
\toprule
Item & Question (short) & Response type & Harmonization rule\\
\midrule
\endhead
\midrule \multicolumn{4}{r}{\emph{Continued on next page}}\\ 
\endfoot
\bottomrule
\endlastfoot
Q1 & Which of the following best describes your current employer? & Categorical (single choice) & Not used / not harmonized in analysis sample\\
Q2 & Who is your current primary employer? & Open text & Not used / not harmonized in analysis sample\\
Q3 & Which of the following best describes your current job? & Categorical (single choice) & Not used / not harmonized in analysis sample\\
Q4 & How many years have you worked for your employer? & Numeric (years/age) & winsorize(1,99) then z-score\\
Q5 & How many hours per week do you work in your job on average? & Numeric (years/age) & winsorize(1,99) then z-score\\
Q6 & What was the estimated income from your job in the past 12 months? & Numeric or ordinal bracket & winsorize(1,99) then log1p then z-score\\
Q7 & Considering your entire household (which includes you, your spouse / partner) now, what was your estimated total household income (including income from all jobs as well as rent, dividends, interest, and other money received) in the past 12 months? & Numeric or ordinal bracket & winsorize(1,99) then log1p then z-score\\
Q8 & What is your age? & Numeric (years/age) & winsorize(1,99) then z-score\\
Q9 & At what age do you plan to retire? & Numeric (years/age) & winsorize(1,99) then z-score\\
Q10 & How would you describe your current health? & Ordinal Likert & z-score of mapped ordinal\\
Q11 & Which of the following, if any, best describes the retirement plans in which you are enrolled through your employer? & Categorical (single choice) & Not used / not harmonized in analysis sample\\
Q12 & What is the name of the hybrid plan in which you are enrolled? & Open text & Not used / not harmonized in analysis sample\\
Q13 & What is the name of the defined benefit pension plan in which you are enrolled? & Open text & Not used / not harmonized in analysis sample\\
Q14 & (To the best of your knowledge,) how much do you expect to receive per year from your defined benefit pension plan after your retirement if you were to leave your job today? & Numeric or ordinal bracket & winsorize(1,99) then log1p then z-score\\
Q15 & For about how many more years do you expect to continue to work for your current employer? & Numeric (years/age) & winsorize(1,99) then z-score\\
Q16 & (To the best of your knowledge,) how much do you expect to receive per year from your defined benefit pension plan after your retirement if you continued to work for the number of years specified in the previous question? & Numeric or ordinal bracket & winsorize(1,99) then log1p then z-score\\
Q17 & To the nearest \$10,000, what is the estimated balance of your current defined contribution plan (e.g. 401(k), 403(b), etc.)? & Numeric or ordinal bracket & winsorize(1,99) then log1p then z-score\\
Q18 & How much do you think your employer pays into your defined benefit pension plan, defined contribution plan, guaranteed return plan and/or hybrid plan combined, as a percentage of your income (before taxes)? & Numeric (percent) & clip to [0,1], logit(eps=1e-4), then z-score\\
Q19 & How much does your household expect to receive annually in retirement benefits after retirement (including all defined benefit plans, 401(k), 403(b), social security benefits, military retired pay and veterans' pensions)? & Numeric or ordinal bracket & winsorize(1,99) then log1p then z-score\\
Q20 & If your employer offered to contribute an amount equal to 2.5\% of your income each year into an investment account, would you enroll in this hypothetical plan if it meant you would stop earning additional benefits under your current plan? & Binary (Yes/No) & Not used / not harmonized in analysis sample\\
Q21 & If your employer offered to contribute an amount equal to 5\% of your income each year into an investment account, would you enroll in this hypothetical plan if it meant you would stop earning additional benefits under your current plan? & Binary (Yes/No) & Not used / not harmonized in analysis sample\\
Q22 & If your employer offered to contribute an amount equal to 7.5\% of your income each year into an investment account, would you enroll in this hypothetical plan if it meant you would stop earning additional benefits under your current plan? & Binary (Yes/No) & Not used / not harmonized in analysis sample\\
Q23 & If your employer offered to contribute an amount equal to 10\% of your income each year into an investment account, would you enroll in this hypothetical plan if it meant you would stop earning additional benefits under your current plan? Table A.1: Pension Survey Questions 48 & Binary (Yes/No) & Not used / not harmonized in analysis sample\\
Q24 & If your employer offered to contribute an amount equal to 15\% of your income each year into an investment account, would you enroll in this hypothetical plan if it meant you would stop earning additional benefits under your current plan? & Binary (Yes/No) & Not used / not harmonized in analysis sample\\
Q25 & If your employer offered to contribute an amount equal to 20\% of your income each year into an investment account, would you enroll in this hypothetical plan if it meant you would stop earning additional benefits under your current plan? & Binary (Yes/No) & Not used / not harmonized in analysis sample\\
Q26 & If your employer offered to contribute an amount equal to 25\% of your income each year into an investment account, would you enroll in this hypothetical plan if it meant you would stop earning additional benefits under your current plan? & Binary (Yes/No) & Not used / not harmonized in analysis sample\\
Q27 & If your employer offered to contribute an amount equal to 30\% of your income each year into an investment account, would you enroll in this hypothetical plan if it meant you would stop earning additional benefits under your current plan? & Binary (Yes/No) & Not used / not harmonized in analysis sample\\
Q28 & If your employer offered to contribute an amount equal to 40\% of your income each year into an investment account, would you enroll in this hypothetical plan if it meant you would stop earning additional benefits under your current plan? & Binary (Yes/No) & Not used / not harmonized in analysis sample\\
Q29 & If your employer offered to contribute an amount equal to 50\% of your income each year into an investment account, would you enroll in this hypothetical plan if it meant you would stop earning additional benefits under your current plan? & Binary (Yes/No) & Not used / not harmonized in analysis sample\\
Q30 & If your employer offered to contribute an amount equal to 60\% of your income each year into an investment account, would you enroll in this hypothetical plan if it meant you would stop earning additional benefits under your current plan? & Binary (Yes/No) & Not used / not harmonized in analysis sample\\
Q31 & Ok, if your employer offered to contribute an amount equal to any percentage of your income each year into an investment account, what is the smallest percentage you would accept to enroll in this hypothetical plan? & Numeric (percent) & Not used / not harmonized in analysis sample\\
Q32 & What healthcare benefits do you expect to receive upon retirement? & Categorical (multi-select) & Not used / not harmonized in analysis sample\\
Q33 & How would you describe the stability of your current retirement plan? & Ordinal Likert & z-score of mapped ordinal\\
Q34 & Suppose you have \$100 in an account with an interest rate of 2\% per year. If you left your money in the account for 5 years, how much money do you think would be in the account? & Financial literacy (correct/incorrect) & z-score of constructed index\\
Q35 & Again, suppose you have \$100 in an account with an interest rate of 2\% per year. If you left your money in the account for 5 years, how much money do you think would be in the account? & Financial literacy (correct/incorrect) & Not used / not harmonized in analysis sample\\
Q36 & Suppose you have some money in an account with an interest rate of 2\% per year. If inflation is 3\%, after one year, will you be able to buy less, more, or exactly the same with the money in your account than you could today? & Financial literacy (correct/incorrect) & Not used / not harmonized in analysis sample\\
Q37 & What typically happens to the value of investment in bonds if interest rates rise? & Financial literacy (correct/incorrect) & Not used / not harmonized in analysis sample\\
Q38 & Buying a single company's stock usually provides a safer return than a stock mutual fund & Financial literacy (correct/incorrect) & Not used / not harmonized in analysis sample\\
Q39 & Suppose you have the option between a secure, guaranteed one-time payment of \$10,000 cash in ten years, or a one-time immediate cash payment today. What is the minimum amount that the immediate cash payment would have to be for you to choose it instead of the payment of \$10,000 in ten years? & Choice (time preference) & Not used / not harmonized in analysis sample\\
Q40 & Given your answer to the previous question, please specify the minimum amount that the immediate cash payment would have to be for you to choose it instead of the payment of \$10,000 in ten years. & Numeric or ordinal bracket & Not used / not harmonized in analysis sample\\
Q41 & What is your sex? & Categorical (single choice) & Not used / not harmonized in analysis sample\\
Q42 & Are you of Hispanic, Latino, or Spanish origin? & Binary (Yes/No) & z-score of binary (optional)\\
Q43 & What is your race? & Categorical (single choice) & Not used / not harmonized in analysis sample\\
Q44 & What is the highest degree or level of school you have completed? & Categorical (single choice) & z-score of mapped education\\
Q45 & What is your marital status? & Categorical (single choice) & Not used / not harmonized in analysis sample\\
Q46 & We thank you for your time spent taking this survey. Is there anything else you'd like to tell us? & Instruction/closing & Not used / not harmonized in analysis sample\\
\end{longtable}
\normalsize

\section{ECV validation protocol and diagnosis}
\label{app:ecv}

This appendix provides operational details for the evaluation and diagnostic procedure summarized in Section~\ref{sec:validation}. The goal is to estimate incremental OOS value while guarding against post-selection bias induced by iterative refinement.

\subsection{Embedded cross-validation (ECV) protocol}
We use an embedded cross-validation design with an \emph{outer} evaluation layer and an \emph{inner} refinement layer.

\paragraph{Outer loop (final evaluation).}
Split the sample into $K_{\text{out}}$ outer folds. For each outer split $k$, reserve fold $k$ as an untouched test set that is never used to choose refinements. All refinement decisions are made using only the remaining data.

\paragraph{Inner loop (refinement and selection).}
Within the outer training data, run $K_{\text{in}}$-fold cross-validation to compute OOS $\Delta$-metrics and diagnostic statistics for each candidate component. Refinement operators (Section~\ref{sec:operators}) may be applied based on inner-loop results. Once a stopping criterion is met, freeze the taxonomy and mapping and evaluate once on the outer test fold.

\paragraph{Standardization and leakage control.}
When evaluating OOS performance, all transformations used to construct scores (e.g., centering/scaling, imputation, any supervised calibration) are fit on the training portion of the relevant fold and applied to the held-out portion. This includes ``train-only'' standardization for respondent-level scores.

\subsection{Incremental OOS value: $\Delta$-metrics and stability}
Let $M(\cdot)$ denote a scalar OOS metric (e.g., $R^2$, RMSE, log loss, AUC) computed on held-out data for a downstream task. For a candidate subdimension $j$, define the incremental gain on fold $f$ as
\[
\Delta M_{j,f} \;=\; M(\text{baseline} + S_j) - M(\text{baseline}),
\]
where $S_j$ is the candidate score and ``baseline'' excludes $S_j$ but may include established covariates and retained scores.
We summarize a candidate by (i) the mean gain $\overline{\Delta M}_j$ and (ii) a stability statistic such as the share of folds with improvement,
\[
p_j \;=\; \frac{1}{K_{\text{in}}}\sum_{f=1}^{K_{\text{in}}}\mathbb{1}\{\Delta M_{j,f} > 0\}.
\]
Classification into \emph{signal/weak-signal/noise-like} uses joint criteria on magnitude and stability; we recommend reporting both $\overline{\Delta M}_j$ and $p_j$ rather than relying on a single threshold.

\subsection{Overlap and misassignment diagnostics}
OOS noise can reflect either insufficient support or overlap-driven contamination. We diagnose overlap using three complementary, lightweight statistics.

\paragraph{(i) Correlation screening.}
Compute pairwise correlations among candidate and retained subdimensions on the training portion of each fold. Persistent high correlations indicate non-separation and motivate cluster-based diagnostics.

\paragraph{(ii) Cross-loading concentration.}
Let $W$ be the item-to-subdimension mapping. For each item, identify whether it loads materially on multiple nearby subdimensions. Concentrated cross-loadings (many items with substantial mass on multiple constructs within a neighborhood) indicate construct contamination. We recommend reporting a simple concentration summary (e.g., the share of items whose top-2 weights are close).

\paragraph{(iii) Conditional contribution within overlap clusters.}
Within each detected overlap cluster, evaluate whether a candidate contributes conditional predictive content beyond its neighbors by comparing a specification that includes the full cluster to one that excludes the candidate. A candidate that fails conditional contribution repeatedly is treated as overlap-driven noise.

\subsection{Data-limitation flags}
We flag a component as \emph{data-limited} when instability is plausibly driven by weak empirical support rather than overlap. Practical indicators include: very low $n_{\text{nonmissing}}$, too few contributing items (effective support), or weak within-sample variation of the score. Data-limited components are reported as coverage constraints and are not routed to refinement.

\subsection{Robustness check}
\label{app:robustness_check}

\subsubsection{(i) Threshold robustness: overlap and sparsification}
We verify that overlap flags and refinement triggers are stable to reasonable threshold choices.
\begin{itemize}[leftmargin=*]
  \item \textbf{Overlap cutoff.} Repeat the correlation screening and overlap-cluster formation under $|\rho|\in\{0.80,0.85,0.90\}$. We report the set of flagged pairs and whether the same ``problem cluster'' appears across cutoffs.
  \item \textbf{Sparsification threshold.} When applying the simplex sparsification step (drop small weights then renormalize), vary the weight threshold $\tau\in\{0.05,0.10,0.15\}$.
  \item \textbf{Top-$m$ constraint.} As an alternative sparsification rule, enforce at most top-$m\in\{1,2,3\}$ nonzero weights per item (keep the top-$m$ weights and renormalize).
\end{itemize}
The goal is not to claim a unique ``correct'' threshold, but to show that the key qualitative conclusions (which constructs overlap; which refinements help) are not driven by a knife-edge parameter choice.

\subsubsection{(ii) Scoring-rule robustness: alternatives to the weighted average}
Our baseline score construction in \eqref{eq:score} is a weighted average across items. We consider three simple alternatives that preserve interpretability and avoid leakage:
\begin{itemize}[leftmargin=*]
  \item \textbf{Weighted sum.} Use $\tilde S_{ik}=\sum_{j} w_{jk}x_{ij}$, optionally followed by within-fold standardization.
  \item \textbf{Item z-scoring before aggregation.} Within each training fold, standardize each item $x_{ij}$ to z-scores (using training-fold means/SDs) and aggregate using the same weights.
  \item \textbf{Missingness-aware reweighting.} Replace $w_{jk}$ with $w'_{jk}\propto w_{jk}\cdot c_j$ where $c_j$ is an item-specific coverage weight (e.g., the training-fold non-missing share), then renormalize so that $\sum_k w'_{jk}=1$ for each item and scores remain comparable.
\end{itemize}
For each alternative, we rerun the same OOS incremental-validity and overlap-diagnostic protocol, comparing (i) the signal/noise classification and (ii) the ranking of the strongest dimensions.

\subsubsection{(iii) Placebo / permutation tests}
We implement two complementary permutation diagnostics.
\begin{itemize}[leftmargin=*]
  \item \textbf{Outcome permutation (null check).} Permute the outcome labels within the evaluation sample and rerun the full OOS incremental-validity pipeline; incremental OOS improvements should concentrate near zero.
  \item \textbf{Mapping permutation (mechanism check).} Randomize the item-to-subdimension weights while preserving key structure (e.g., per-item simplex and the same sparsity pattern), then rerun the OOS evaluation. If the original LLM mapping captures meaningful semantic structure, its OOS deltas should dominate the distribution generated by randomized mappings.
\end{itemize}
These tests are designed to be ``highly diagnostic'' rather than to optimize performance: they help rule out mechanical OOS gains from adding arbitrary predictors or from overfitting the mapping.

\subsection{Decision logic and reporting}
Each inner-loop iteration produces (a) $\overline{\Delta M}_j$ and $p_j$ for OOS value and (b) overlap diagnostics and data-support flags. We retain components with stable OOS gains and adequate separation (or that pass conditional contribution within clusters). Noise-like components are either deferred as data-limited or refined when overlap diagnostics indicate contamination. Components that repeatedly fail triage and are not improved by targeted refinements are discarded. For transparency, we recommend reporting a triage table listing $\overline{\Delta M}_j$, $p_j$, overlap indicators, and the resulting action.

\subsection{Refined component definitions and weight shares}\label{app:split_shares}
Table~\ref{app:tab:generosity_split_allocation} consolidates (i) the v1 cross-loadings that drive the Financial Literacy--Generosity overlap and (ii) the implied allocation of v1 ``perceived generosity'' weight mass across the refined v2 child subdimensions.

\begin{table}[!htbp]\centering
\caption{From overlap to refinement: v2 reallocation across dimensions and refined belief channels}\label{app:tab:generosity_split_allocation}
\footnotesize
\begin{threeparttable}
\begin{adjustbox}{max width=0.92\textwidth}
\begin{tabular}{lrrrr}
\toprule
Item & Financial Literacy & Generosity: benefit value & Generosity: employer contribution & Plan stability\\
\midrule
Q14 & 0.25 & 0.75 & 0.00 & 0.00 \\
Q16 & 0.25 & 0.75 & 0.00 & 0.00 \\
Q18 & 0.40 & 0.00 & 0.60 & 0.00 \\
Q33 & 0.05 & 0.00 & 0.00 & 0.95 \\
\bottomrule
\end{tabular}
\end{adjustbox}
\begin{tablenotes}[flushleft]\footnotesize
\item Notes: The Financial Literacy column corresponds to the \texttt{financial\_literacy} dimension. The two ``Generosity:'' columns are child subdimensions created by splitting the parent \texttt{perceived\_generosity} subdimension within DB beliefs (Q14/Q16 $\to$ benefit value; Q18 $\to$ employer contribution). ``Plan stability'' (Q33) is treated as a separate belief channel and reported as its own column to align with the incremental OOS tests in Appendix Table~\ref{app:tab:oos_v2_generosity_split}.
\end{tablenotes}
\end{threeparttable}
\end{table}

\begin{table}[!htbp]\centering
\caption{OOS incremental validity for the v2 local refinement of perceived generosity}\label{app:tab:oos_v2_generosity_split}
\footnotesize
\begin{adjustbox}{max width=\textwidth}
\begin{tabular}{lccc ccc}
\toprule
& \multicolumn{3}{c}{Acceptance (binary)} & \multicolumn{3}{c}{Contribution rate $\,|\,accept{=}1$} \\
\cmidrule(lr){2-4}\cmidrule(lr){5-7}
Model addition & $\Delta$AUC (mean) & $\Delta$AUC (median) & Share improve & $\Delta R^2$ (mean) & $\Delta R^2$ (median) & Share improve \\
\midrule
Add generosity: benefit value (Q14,Q16) & 0.0037 & 0.0055 & 0.76 & 0.0016 & 0.0017 & 0.84 \\
Add generosity: employer contribution (Q18) & -0.0006 & 0.0009 & 0.60 & -0.0004 & 0.0003 & 0.56 \\
Add generosity: both & 0.0033 & 0.0052 & 0.72 & 0.0012 & 0.0018 & 0.80 \\
Incremental: add plan stability (Q33) on top of both & 0.0193 & 0.0204 & 0.84 & 0.0002 & 0.0002 & 0.64 \\
\bottomrule
\end{tabular}
\end{adjustbox}
\begin{minipage}{0.92\linewidth}\footnotesize
\emph{Notes:} All numbers are out-of-sample (OOS) improvements relative to the same baseline specification (with the canonical Financial Literacy battery anchored). ``Share improve'' is the fraction of repeated splits in which the augmented model improves the corresponding OOS metric. For contribution rates, the sample is restricted to accepters.
\end{minipage}
\end{table}

\begin{figure}[!htbp]\centering
\caption{LLM refinement prompt (anchor-then-split) used to reduce cross-loading between Financial Literacy and perceived generosity}\label{app:fig:prompt_refinement}
\footnotesize
\begin{minipage}{0.95\textwidth}
\begin{verbatim}
Prompt template (copy/paste to generate v2 codebook)


You are refining a survey item taxonomy for an econometric measurement paper.
Task: For each survey question, assign (i) one PRIMARY subdimension (weight=1.0) and optionally (ii) one SECONDARY subdimension (weight between 0.05 and 0.20).
You must apply the definitions and boundary rules below and avoid label confusion between financial_literacy and perceived_generosity.

Definitions:
- Cognition_time.financial_literacy: objective knowledge/ability about finance concepts/calculations/tradeoffs.
- DB_beliefs.perceived_generosity: perceived generosity/attractiveness of employer plan (match/benefits) and evaluative judgments.
- NEW: Controls.plan_awareness: awareness/knowledge of plan rules/features (match rate, contribution rules, vesting) without evaluative generosity language.

Constraints:
- Exactly 1 primary label per item (weight 1.0).
- At most 1 secondary label (0.05–0.20). If none, omit.
- Provide: (a) one-sentence rationale; (b) one-sentence "why not" addressing the closest alternative label.

Return results in a CSV-like table with columns:
Q, question_text, primary_label, secondary_label(optional), notes

Now label these items:
{ITEM_LIST}
\end{verbatim}
\end{minipage}
\end{figure}

\subsection{Robustness checks}
\label{app:robustness_suite}

This subsection records robustness exercises that accompany the main workflow in Section~\ref{sec:workflow}. They are intentionally lightweight: each exercise can be implemented by rerunning the same embedded cross-validation (ECV) pipeline under a small grid of alternative settings.

\begin{table}[H]
\centering
\caption{Robustness to alternative score-construction rules for refined belief channels}
\label{tab:robust_scoring_rules}
\footnotesize
\setlength{\tabcolsep}{6pt}
\renewcommand{\arraystretch}{1.15}

\begin{threeparttable}
\begin{adjustbox}{max width=0.95\linewidth}
\begin{tabular}{p{5.4cm}rr rr}
\toprule
& \multicolumn{2}{c}{Add benefit + employer} 
& \multicolumn{2}{c}{Add + stability} \\
\cmidrule(lr){2-3}\cmidrule(lr){4-5}
Scoring rule 
& $\Delta$AUC 
& $\Delta R^2$ 
& $\Delta$AUC 
& $\Delta R^2$ \\
\midrule
Use tilde (current)        
& 0.0007  & 0.0007  & 0.0066  & 0.0003 \\
Re-zscore items then mean  
& 0.0007  & 0.0007  & 0.0066  & 0.0003 \\
Weighted sum (Q14+Q16)     
& $-0.0003$ & 0.0002  & 0.0091  & $-0.0014$ \\
\bottomrule
\end{tabular}
\end{adjustbox}

\begin{tablenotes}[flushleft]\footnotesize
\item \emph{Notes:} ``Use tilde'' uses harmonized item scores as constructed in Appendix~\ref{app:harmonization}.
``Re-zscore'' standardizes items before aggregation (within-fold).
``Weighted sum'' replaces the mean with a sum for the benefit-value channel (Q14 and Q16).
Reported out-of-sample deltas are relative to the same baseline specification.
\end{tablenotes}
\end{threeparttable}
\end{table}

\begin{table}[H]\centering
\caption{Robustness to sparsification parameters in the initial soft mapping (illustration: top-$m=2$).}\label{tab:robust_tau_topm}
\footnotesize
\begin{threeparttable}
\begin{adjustbox}{max width=0.92\linewidth}
\begin{tabular}{lrrrr}
\toprule
$\tau$ & $\Delta$AUC (accept) & $\Delta R^2$ (rate$\,|\,\text{accept}=1$) & $\rho(\text{FinLit},\text{Generosity})$ & \#pairs $|\rho|\ge 0.80$ \\
\midrule
0.05 & -0.0037 & 0.0009 & 0.444 & 1 \\
0.10 & 0.0006 & 0.0005 & 0.440 & 1 \\
0.15 & -0.0033 & 0.0017 & 0.443 & 0 \\
\bottomrule
\end{tabular}
\end{adjustbox}
\begin{tablenotes}[flushleft]\footnotesize
\item Notes: Each row re-runs the same out-of-sample protocol after sparsifying each item-to-subdimension weight vector by dropping weights below $\tau$ and renormalizing the retained mass (here with an additional top-$m=2$ cap). The last two columns summarize overlap diagnostics; the correlation is computed using respondent-level subdimension scores constructed under the corresponding setting.
\end{tablenotes}
\end{threeparttable}
\end{table}

\begin{table}[H]\centering
\caption{Permutation placebo for incremental validity: randomized mapping vs. observed mapping.}\label{tab:robust_placebo}
\footnotesize
\begin{threeparttable}
\begin{adjustbox}{max width=0.92\linewidth}
\begin{tabular}{lrr}
\toprule
 & $\Delta$AUC (accept) & $\Delta R^2$ (rate$\,|\,\text{accept}=1$) \\
\midrule
Observed mapping & -0.0003 & 0.0016 \\
Placebo mean & -0.0011 & 0.0007 \\
Placebo sd & 0.0012 & 0.0012 \\
Placebo min & -0.0032 & -0.0015 \\
Placebo max & 0.0013 & 0.0032 \\
$p$-value (share placebo $\ge$ observed) & 0.30 & 0.20 \\
\bottomrule
\end{tabular}
\end{adjustbox}
\begin{tablenotes}[flushleft]\footnotesize
\item Notes: Each placebo draw preserves the per-item sparsity pattern and weight magnitudes but randomly permutes the subdimension labels within item (destroying semantic alignment while preserving mechanical complexity). We then reconstruct subdimension scores and re-run the identical OOS evaluation.
\end{tablenotes}
\end{threeparttable}
\end{table}

\subsection{Triage summary table}
\label{app:triage}

\begin{table}[H]
\centering
\caption{OOS triage summary: signal versus noise and refinement targets}
\label{app:tab:triage-summary}

\footnotesize
\setlength{\tabcolsep}{3pt} 
\renewcommand{\arraystretch}{1.2}

\begin{tabularx}{\textwidth}{
  l 
  l 
  >{\raggedright\arraybackslash}X 
  r 
  r 
  l 
  l 
  >{\raggedright\arraybackslash}X 
}
\toprule
Family & Subdimension & Items & $\Delta$AUC & $\Delta R^2$ & Accept & Rate & Notes \\
\midrule

Cognition\_time & financial\_literacy & Q14, Q16, Q17, Q18, Q33 & 0.0817 & 0.0009 & Strong & Weak & \\

Controls & demographics & Q42, Q44, Q7, Q8 & 0.0189 & $-0.0004$ & Weak & Noise & \\

Controls & employment\_context & Q4, Q5, Q6 & $-0.0022$ & 0.0006 & Noise & Weak & \\

DB\_beliefs & generosity: benefit value & Q14, Q16 & 0.0037 & 0.0016 & Weak & Weak & v2 split \\

DB\_beliefs & generosity: benefit + contrib. & Q14, Q16, Q18 & 0.0033 & 0.012 & Weak & Weak & v2 split \\

DB\_beliefs & generosity: employer contrib. & Q18 & $-0.0006$ & $-0.0004$ & Noise & Noise & v2 split \\

DB\_beliefs & perceived\_generosity & Q14, Q16, Q18 & $-0.0026$ & $-0.0005$ & Noise & Noise & \\

DB\_beliefs & plan stability (incremental) & Q33 & 0.0193 & 0.0002 & Strong & Weak & Inc. over v2 split \\

Econ\_constraints & income\_wealth\_buffer & Q17, Q19, Q6, Q7 & 0.0358 & $-0.0004$ & Weak & Weak & \\

Econ\_constraints & retirement\_horizon & Q15, Q19, Q9 & 0.0239 & 0.0003 & Weak & Weak & \\

Econ\_constraints & service\_tenure\_lockin & Q15, Q4 & 0.1143 & 0.0089 & Strong & Medium & \\

\bottomrule
\end{tabularx}

\vspace{0.8em}
\begin{minipage}{\textwidth}
    \footnotesize
    \emph{Notes:} $\Delta$AUC is the out-of-sample improvement in AUC for acceptance relative to the baseline model. $\Delta R^2$ is the out-of-sample improvement in $R^2$ for the contribution-rate regression estimated on the acceptor subsample. ``Plan stability (incremental)'' reports the additional gain from adding plan stability on top of the v2 generosity split.
\end{minipage}
\end{table}

\end{document}